# Well-being policy evaluation methodology based on WE pluralism

## Takeshi Kato

### Abstract

Methodologies for evaluating and selecting policies that contribute to the well-being of diverse populations need clarification. To bridge the gap between objective indicators and policies related to well-being, this study shifts from constitutive pluralism based on objective indicators to conceptual pluralism that emphasizes subjective context, develops from subject–object pluralism through individual–group pluralism to WE pluralism, and presents a new policy evaluation method that combines joint fact-finding based on policy plurality. First, to evaluate policies involving diverse stakeholders, I develop from individual subjectivity–objectivity to individual subjectivity and group intersubjectivity, and then move to a narrow–wide WE pluralism in the gradation of I–family–community–municipality–nation–world. Additionally, by referring to some functional forms of well-being, I formulate the dependence of well-being on narrow–wide WE. Finally, given that policies themselves have a plurality of social, ecological, and economic values, I define a set of policies for each of the narrow–wide WE and consider a mapping between the two to provide an evaluation basis. Furthermore, by combining well-being and joint fact-finding on the narrow–wide WE consensus, the policy evaluation method is formulated. The fact–value combined parameter system, combined policy-making approach, and combined impact evaluation are disclosed as examples of implementation. This paper contributes to the realization of a well-being society by bridging philosophical theory and policies based on WE pluralism and presenting a new method of policy evaluation based on subjective context and consensus building.

# 1 Introduction

**Policies** that contribute to **well-being** are needed worldwide [1,2]. The Organisation for Economic Cooperation and Development (OECD) and the Wellbeing Economy Alliance (WEAll) present a well-being-oriented economy [3,4]. The United Nations Sustainable Development Goals [5] include "health and well-being for all" as Goal 3. Japan's Digital Garden City Nation Initiative [6] and Smart Wellness City Initiative [7] aim to improve well-being. In the early stages of economic development, economic indicators such as production and income were important because the satisfaction of basic needs was the main issue. However, as basic needs are satisfied, it has become necessary to consider well-being related to non-economic factors such as life feelings and social relationships [1].

Various indicators have been developed to evaluate policies that contribute to well-being, including the UN's Human Development Index, the OECD's Better Life Index, and WEAll's Happy Planet Index [8]. However, in the Well-being Indicators that comprise the OECD's Better Life Index, objective indicators for income, housing, health, education, environment, and safety do not necessarily correspond to nation rankings of subjective life satisfaction [9]. The Liveable Well-Being City Indicators [10], developed by Japan for its Digital Garden City Nation Initiative, found little correlation between 22 categories of objective indicators and 33 categories of subjective indicators for 1021 municipalities [11]. In other words, there is a gap between subjective and objective indicators, and it is necessary to reexamine how policies should be evaluated.

Mitchell et al. discuss the science and philosophy of well-being [12]. The science of well-being includes **measurement pluralism**: a position that seeks to measure constructs of well-being by selecting a variety of indicators, and **methodological pluralism**: a position that seeks to parallel or combine various approaches such as economics, sociology, and psychology, as well as construct selection and measures. Measurement pluralism argues that the Well-being Indicators and Liveable Well-Being City Indicators described above cannot be a generic gold standard, and methodological pluralism requires not only a diverse selection of indicators but also a combination of various quantitative and qualitative methodologies at the psychological, cultural, and social levels.

Compared with these sciences, philosophies of well-being include **constitutive pluralism**: a position that well-being is composed of multiple elements that are not reducible to each other, and **conceptual pluralism**: a position that the appropriate concept of well-being depends on the **subjective and social context** of people's facts, environment, and goals. Constitutive pluralism attempts to explain well-being comprehensively based on a list of goods or indicators, whereas conceptual pluralism assumes that the constructs differ from person to person or from period to period of life. If a constitutivist modify constitutive pluralism to take the position that different constructs are involved according to diverse contexts, it is little different from conceptual pluralism. Mitchell et al. then argue that the methodological pluralism of science recognizes the constructive concepts of philosophy, while the conceptual pluralism of philosophy must take on the methodology of science for social practice and combine the two.

Even limiting the constructs of well-being to subject and object for simplicity, according to Ishida, there are **subjectivism**, **objectivism**, and **subject–object pluralism** [13]. Subject–object pluralism states that a hybrid of individual's subjective attitudes and objective values that are independent of them is conducive to well-being. This can be described as a hybrid of constitutive pluralism and conceptual pluralism. It is further classified into symmetrical and asymmetrical pluralism depending on whether subjectivity and objectivity are treated symmetrically or not. In symmetric pluralism, split cases may occur where subjective and objective evaluations are divided. But in the asymmetric pluralism proposed by Ishida, subjective attitude is a monotonically increasing function with positive and negative values, objective value is a function with positive values, and well-being is formulated as the product of both functions to avoid split cases.

According to Mitchell et al. and Ishida's classification, the OECD's Well-being Indicators and Japan's Liveable Well-Being City Indicators, mentioned above, belong to measurement pluralism and methodological pluralism in the scientific sense, and constitutive pluralism and subject–object pluralism in the philosophical sense, in that they compare the well-being of multiple countries and municipalities using multiple objective and subjective indicators and findings from economics, sociology, and psychology. In these indicators, the context of people in conceptual pluralism is either not taken into account or it is implicitly assumed that the users of the indicators use them with context in mind.

Objective indicators have a certain significance in comparing the sufficiency of material conditions and social infrastructure, but the problem is that, as already mentioned, there is a gap between objective and subjective indicators. Policies involve the investment of objective goods such as funds and personnel, but even if policies are implemented based on objective indicators, they do not necessarily lead to subjective well-being. In other words, in order to implement policies that contribute to well-being, policies need to be pre-evaluated and selected based on indicators linked to subjectivity. This is a position that leans toward subjectivism and, if context is taken into account, a shift from constitutive pluralism to conceptual pluralism.

Another problem is that policy involves **consensus building** not by a single individual but by a group consisting of many stakeholders, and on top of that, fairness, rationality with respect to situation analysis and future projections, efficiency with respect to cost and time, and stability after agreement are also required [14]. In other words, even if policies are evaluated based on indicators linked to subjectivity, it is necessary to fairly take diverse subjectivities into account, which leads to a new consideration of **individual–group pluralism**. Moreover, given that the policy itself encompasses a variety of elements such as funds, workforce, time, effects, and side effects, it is imperative to consider the **policy plurality**. The combination of parameters among the constructive elements will generate a plethora of policy proposals. It is difficult to select a policy from among them based solely on subjectivity; therefore, an objective indicator must be introduced once more. Furthermore, if rationality and efficiency are sought to convince a large number of stakeholders, **joint fact-finding** is required [15]. Therefore, it is necessary to evaluate policies that contribute to well-being based on consensus by incorporating not only subjective indicators but also objective joint facts again, and combining that with joint fact-finding while respecting subjective context.

The purpose of this paper is to bridge the gap between indicators and policies on well-being and provide a methodology for evaluating policies conducive to well-being. To this end, I first take the positions of methodological pluralism and subject–object pluralism. I derive individual–group pluralism by replacing objectivity with group intersubjectivity to incorporate conceptual pluralism (subjective context). I then further develop individual subjectivity and group intersubjectivity into a new **WE pluralism** based on the view of "self as WE." I then attempt to formulate a methodology for evaluating well-being policies by reintroducing intersubjective and objective joint fact-finding into WE pluralism while taking into account policy plurality in addition to WE plurality.

This paper proceeds as follows. In Chapter 2, I first refer to Ishida's subject–object pluralism formula and expand it through individual–group pluralism to WE pluralism, which incorporates a conceptual pluralism that leans toward subjectivism. In Chapter 3, I extend WE pluralism by incorporating policy plurality. I formulate a consensus in "narrow–wide WE" and present a method for evaluating policies by combining consensus building and joint fact-finding. Then, I disclose three practical examples based on this policy evaluation method. Finally, in Chapter 4, I summarize the conclusions of this paper and discuss future issues.

## 2 WE pluralism

### 2.1 From subject–object to individual–group pluralism

First, I refer to subject–object pluralism. This is the position that subjectivity and objectivity are independent of each other, and that a hybrid of an individual $p$'s subjective attitude for policy $x$ and objective values independent of it contributes to well-being. Based on Woodard's classification [16], Ishida expresses subject–object pluralism as follows [13].

**Subject–object pluralism**: (1) If an individual $p$ has a positive attitude for $x$ and $x$ has goodness independent of the $p$'s positive attitude, then $x$ contributes to the $p$'s well-being. (2) If an $p$ has a negative attitude for $x$ and $x$ has badness independent of the $p$'s negative attitude, then $x$ impairs the $p$'s well-being.

Let $W_s(x, p)$ be the subjective attitude of $p$ for $x$ and $W_o(x)$ be the objective value independent of it. Then, well-being $W(x, p)$ can be expressed as the simplest expression as follows [13].

$$W(x,p) = W_s(x,p) + W_o(x) \tag{1}$$

I would now like to reconsider objective value, which does not depend on subjective attitudes. As discussed in Chapter 1, indicators of objective value are not necessarily tied to subjective well-being. Let me also reconsider that objective value is calculated from some data. According to Deguchi, even the physical constants that can be viewed as the most objective come from a contest of certainty that combines various physical measurements and statistical treatments, with no basis for the assumptions underlying the statistical processing, which were accepted as the result of a collaborative effort [17,18]. According to Otsuka, in data processing, there is an assumption of the ontology of the world of probability models in its premises. There are different semantic interpretations of the probability model, subjectivism and frequentism, and the decisions that justify the results are based on epistemological beliefs [19].

Given these considerations, an objective value is a value that is determined to be correct through the joint work and interpretation of a large number of people, which can also be thought of as the intersubjectivity of a group. In other words, subjectivity and objectivity are positioned as individual subjectivity and intersubjectivity of a larger group, and subject–object pluralism is replaced by individual–group pluralism.

**Individual–group pluralism**: If an individual $p$ has subjective value for $x$ and a group $g$ has intersubjective value for $x$, then $x$ contributes to the well-being of $p$ and $g$. (Disjunction regarding negative values is omitted.)

Replacing the objective value $W_o(x)$ in Equation (1) by the intersubjectivity $W_{is}(x,g)$ of group $g$, well-being $W(x)$ in individual–group pluralism can be expressed as follows.

$$W(x) = W_s(x,p) + W_{is}(x,g) \tag{2}$$

Here, $W_s(x,p)$ and $W_{is}(x,g)$ do not necessarily have to be composed of the same elements or methodologies. Even if they are the same elements, they can be discarded in consideration of the context of individual $p$ and group $g$. From this perspective, Equation (2) can be viewed as methodological pluralism and conceptual pluralism. In addition, the fact that the subjectivity of individual $p$ and the intersubjectivity of group $g$ differ is a situation that can actually occur, and I do not dare here to formulate a formulation that avoids split cases between the two.

## 2.2 From individual–group to WE pluralism

Next, I will examine the individual and the group. Deguchi describes the "self as WE," an East Asian view of the self that is connected to the lineage of Lao Zhuang and Zen thought, as opposed to the Western view of the "self as I" [20,21]. The "self as WE" is a multi-agent network system of entrustment that includes "I" and other people and things around "I."

From this perspective, the dichotomies of subjectivity–objectivity and individual–group no longer exist, and these are positioned relative to the gradations between, for example, I–family–community–municipality–nation–world. Individual–group pluralism is replaced by "WE pluralism" as the relative openness and spread of WE, and for the sake of simplicity, the two can be taken out as a "narrow–wide WE" pluralism. In Deguchi's words, this can be called the "WE turn" of individual–group pluralism.

> **WE pluralism**: If a "narrow WE" has subjective value for $x$, and a "wide WE" has subjective value for $x$, then $x$ contributes to the well-being of "narrow WE" and "wide WE." (Disjunction regarding negative values is omitted.)

Replacing the subjective value $W_s(x, p)$ of individual $p$ and the intersubjective value $W_{is}(x, g)$ of group $g$ in Equation (2) by the subjective value $W_n(x)$ of "narrow WE" and $W_w(x)$ of "wide WE," respectively, WE pluralism is expressed as follows.

$$W(x) = W_n(x) + W_w(x) \qquad (3)$$

Note that $W_n(x)$ and $W_w(x)$ do not necessarily consist of the same elements as described in the individual–group pluralism, but include the context of conceptual pluralism. Even if the "narrow WE" is included within the "wide WE," just as the subjective value $W_s(x, p)$ of individual $p$ and the intersubjective value $W_{is}(x, g)$ of group $g$ may differ, the subjective value $W_n(x)$ that the "narrow WE" has for $x$ in its context and the subjective value $W_w(x)$ that the "wide WE" has for $x$ in its context may differ, so each is taken into account to represent the overall well-being $W(x)$. Also, both are only two representatives taken out of the spread of WEs, and by considering a variety of WEs, Equation (3) can be expanded into a polynomial equation. Each term is a function with positive or negative values, and the overall well-being $W(x)$ will vary depending on what weights are used to add up the value of policy $x$ for each WE.

I now consider specific functional forms of $W_n(x)$ and $W_w(x)$ in order to picture how $W(x)$ changes. Figure 1 shows three graphs drawing typical functional forms proposed for well-being.

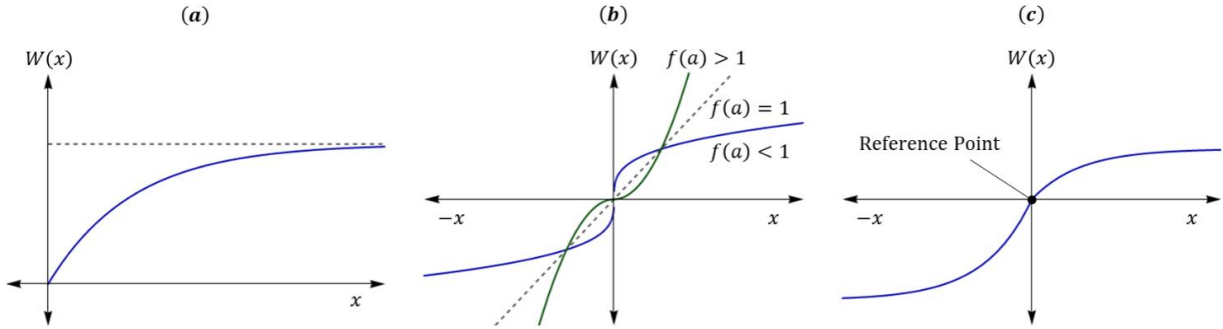

Figure 1. (a) Kagan, (b) Sarch, and (c) prospect theory.

Kagan uses a saturated form of the function to represent the gradually decreasing increment of well-being $W(x)$ relative to the increment of pleasure $x$ for objective goods, as shown in Figure 1 (a) [22]. Sarch expresses the discounting/inflating of the increment of well-being $W(x)$ relative to the increment of enjoyment of life, as shown in Figure 1 (b), with the net enjoyment $x$ at base and the conditional function $f(a)$ of achievement $a$ as power exponent as follows [23]. Whether it is in discounted or inflated form depends on whether $f(a)$ exceeds the threshold value of 1 or not.

$$W(x) = \begin{cases} x^{f(a)} & if \ x \geq 0 \\ -|x|^{f(a)} & if \ x < 0 \end{cases} \tag{4}$$

As functional forms of well-being, the utility function in economics and the value function in behavioral economics prospect theory may be helpful. A utility function is a function of the magnitude of utility relative to preference. By reading the preference as a positive attitude or subjective value for $x$ and the utility as well-being, the utility function can be referred as a function of well-being. In prospect theory, the value function is multiplied by a probability weighting function to represent expected utility, while the value function corresponds to the utility function in general economics. Prospect theory uses an asymmetric functional form in which the value function is expressed in saturated or discounted form, and in which ill-being is perceived as greater than well-being, as shown in Figure 1 (c). Specifically, several functions have been proposed as follows [24].

$$W(x) = \begin{cases} x & \text{Linear} \\ \ln(a+x) & \text{Logarithmic} \\ x^a & \text{Power} \\ ax - x^2 & \text{Quadratic} \\ 1 - e^{-ax} & \text{Exponential} \\ bx - e^{-ax} & \text{Lin.+Exp.} \end{cases} \quad if \ x \geq 0 \tag{5}$$

Among these functions, the Power form (corresponding to Sarch) or the Exponential form (corresponding to Kagan) is generally used [24,25]. Here, given that the function of well-being with respect to income (a preference) is in saturated form [1], and using the exponential form following Kagan, the asymmetric function for positive and negative $x$ is expressed as follows.

$$W(x) = \begin{cases} 1 - e^{-\alpha x} & if \ x \geq 0 \\ -\lambda\left(1 - e^{\beta x}\right) & if \ x < 0 \end{cases} \tag{6}$$

Considering the different context and subjective value of "narrow WE" and "wide WE" as WE pluralism, $W_n(x)$ and $W_w(x)$ in Equation (3) are expressed as follows: $x_n$ and $x_w$ are variables, $\alpha_n, \beta_n, \lambda_n, \alpha_w, \beta_w$ and $\lambda_w$ are coefficients, and $r \ (0 \leq r \leq 1)$ is the weight for normalization.

$$W = r \cdot W_n(x_n) + (1-r) \cdot W_w(x_w) \tag{7}$$

$$W_n(x_n) = \begin{cases} 1 - e^{-\alpha_n x_n} & if \ x_n \geq 0 \\ -\lambda_n\left(1 - e^{\beta_n x_n}\right) & if \ x_n < 0 \end{cases}$$
$$W_w(x_w) = \begin{cases} 1 - e^{-\alpha_w x_w} & if \ x_w \geq 0 \\ -\lambda_w\left(1 - e^{\beta_w x_w}\right) & if \ x_w < 0 \end{cases} \tag{8}$$

Thus, the variation of $W$ with respect to the "narrow–wide WE" can be seen. $W_n(x_n)$ and $W_w(x_w)$ are each composed of several elements, but it is assumed that they are aggregated into a single variable $x_n$ and $x_w$, respectively. For simplicity, the coefficients of both are $\alpha_n = \beta_n = \alpha_w = \beta_w = 1$ and $\lambda_n = \lambda_w = 2$. The reason for setting $\lambda_n = \lambda_w = 2$ is to represent asymmetry between well-being and ill-being. Figure 2 shows a graph of the calculation results.

In Figure 2, the graph asymptotes to $W \to 1$ when $x_n \to \infty$ and $x_w \to \infty$, and to $W \to -2$ when $x_n \to -\infty$ and $x_w \to -\infty$. As shown in Figure 2 (a)–(c), the shape of the 3-dimensional surface changes depending on whether "narrow WE" or "wide WE" is weighted (value of $r$). That is, in (a), well-being is determined by both "narrow WE" and "wide WE," while in (b) and (c) it is determined almost exclusively by one or the other. The (d) shows the case where both are equal, i.e., both reach a consensus, and the change in well-being is represented by a curve rather than a 3-dimensional surface.

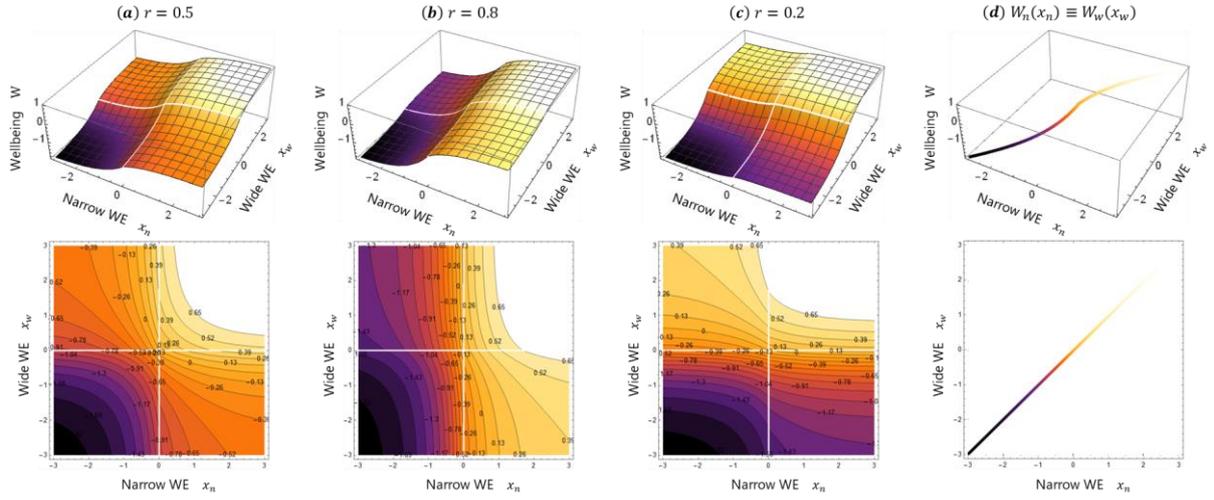

Figure 2. Well-being calculation results for "narrow–wide WE:" (a) $r = 0.5$ ("narrow WE" and "wide WE" are symmetrical), (b) $r = 0.8$ (weighting "narrow WE"), (c) $r = 0.2$ (weighting "wide WE"), and (d) $W_n(x_n) \equiv W_w(x_w)$ ("narrow WE" and "wide WE" are equal).

## 3 Policy evaluation method

### 3.1 Policy evaluation based on policy plurality and WE consensus

From the standpoint of conceptual pluralism in WE pluralism, the "narrow WE" $W_n(x_n)$ and "wide WE" $W_w(x_w)$ elements for a given policy are pluralistic and differ for each context.

> **Case A**: Policies to introduce natural energy may contribute to wellbeing, but this does not necessarily coincide with reduced energy costs, reduced carbon footprint, protection of natural landscapes, increased local economic circulation, etc. [26], and the context varies by the spread of WE: I–family–community –municipality–nation–world.
>
> **Case B**: Policies to enhance social security may contribute to the well-being of the covered population, but this may lead to higher taxes and administrative costs, suppression of other budgets and personnel (education, industrial development, environmental protection, etc.), distortion of social justice, and reduced work incentives [27].

Taking into account the policy plurality and context, Equation (7) can be rewritten and expressed as follows: Let $X_n$ and $X_w$ be the sets of elements and $\boldsymbol{x_n}$ and $\boldsymbol{x_w}$ be vectors consisting of elements with respect to "narrow WE" and "wide WE."

$$W = r \cdot W_n(\boldsymbol{x_n}) + (1-r) \cdot W_w(\boldsymbol{x_w}) \tag{9}$$

$$X_n = \{x_{n1},\, x_{n2},\, \cdots,\, x_{ni},\, \cdots\}$$
$$X_w = \{x_{w1},\, x_{w2},\, \cdots,\, x_{wj},\, \cdots\}$$

Here, I assume that there is a relationship between the components of the "narrow WE" and the "wide WE" for a given policy. As shown in Figure 3 (a), given the mapping $f$ from set $X_w$ to set $X_n$ (the inverse mapping $f^{-1}$ from set $X_n$ to set $X_w$), Equation (9) can be expressed using the common variable vector $\boldsymbol{x_w}$ as follows.

$$\boldsymbol{x_n} = f(\boldsymbol{x_w})$$

$$W = r \cdot W_n \circ f(\boldsymbol{x_w}) + (1 - r) \cdot W_w(\boldsymbol{x_w})$$

(10)

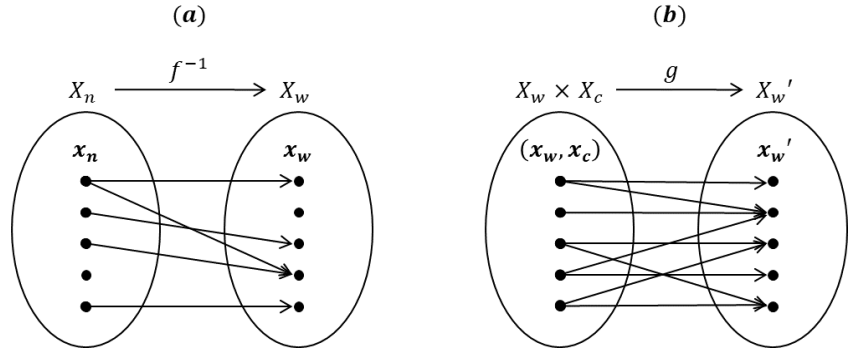

Figure 3. Mapping of well-being element sets: (a) $X_n \xrightarrow{f^{-1}} X_w$ and (b) $X_w \times X_c \xrightarrow{g} X_w{}'$.

Furthermore, if there is consensus between "narrow WE" and "wide WE" for a given policy, i.e., if their contexts match, then they both have the same function for the variable vector $\boldsymbol{x_w}$ and their weights $r$ will disappear. Equation (10) can be rewritten as the following concise function, which corresponds to Figure 2 (d).

$$W_n \circ f \equiv W_w$$

$$W = W_w(\boldsymbol{x_w})$$

(11)

Now, I have moved from subject–object pluralism, individual subjectivity–group intersubjectivity pluralism, and "narrow–wide WE" pluralism, policy and context plurality (conceptual pluralism), to "narrow–wide WE" consensus. However, as discussed in Chapter 1, consensus building requires rationality, efficiency, and stability along with fairness, and joint fact-finding is required [14,15]. As discussed in Chapter 1, it is difficult for people to judge the vast number of policy proposals resulting from the combination of elements solely on the basis of their subjectivity, and as demonstrated in Case A and Case B, it is impossible for people to grasp the complex relationships among the various elements solely on the basis of their subjectivity. While Section 2.1 mentioned that objective values are subjective values determined to be correct through people's collaboration and interpretation, it is useful to reiterate the intersubjectivity that all of the various stakeholders would be able to accept as correct, again as objective joint facts.

Therefore, rather than replacing "narrow–wide WE" consensus for joint facts, I assume that the joint facts perturbatively affect the consensus. As shown in Figure 3 (b), considering the mapping g from the direct product $X_w \times X_c$ of the common set $X_w$ of the "narrow–wide WE" and the set $X_c$ of joint facts to the set $X_w{'}$ incorporating the influence of $X_c$, Equation (11) can be rewritten using vectors $\boldsymbol{x_w}, \boldsymbol{x_c}$ and $\boldsymbol{x_w{'}}$ as follows.

$$X_c = \{x_{c1}, \ x_{c2}, \ \cdots, \ x_{ck}, \ \cdots \} \tag{12}$$

$$\boldsymbol{x_w{'}} = g(\boldsymbol{x_w}, \boldsymbol{x_c})$$
$$W' = W_w(\boldsymbol{x_w{'}}) \tag{13}$$

The meaning of Equation (13) is not to enumerate a list of objective indicators, as is the case with constitutive pluralism, but to explain well-being by combining joint fact-finding for policies into it, while emphasizing the subjective context in conceptual pluralism. In other words, as shown in Figure 4, well-being $W' = W_w(\boldsymbol{x_w{'}})$ is expressed with the agreed subjective value $\boldsymbol{x_w}$ as the main term and the objective indicators based on joint fact-finding as the perturbation term $\boldsymbol{x_c}$. In this way, it is possible to assess the effects of policies on well-being under a "narrow–wide WE" consensus, while avoiding the gap between subjective and objective indicators as described in Chapter 1.

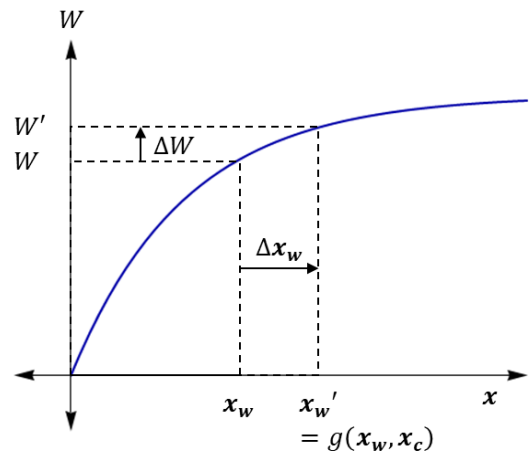

Figure 4. Well-being policy evaluation method based on joint fact-finding under WE consensus.

Note that Equations (9)–(13) were expressed in terms of "narrow WE" and "wide WE," but when a variety of WEs are involved, they can be expanded to polynomials and formulated using the same approach as before. Although Equation (11) shows the case where there is only one consensus, it is also possible, for example, to evaluate policies for the consensus of each stakeholder group and compare them to consider better well-being policies for the whole.

## 3.2   Practical examples of policy evaluation method

This section discloses a method for evaluating policies under WE consensus based on Equations (9)–(13). Figure 5 illustrates the three practical examples.

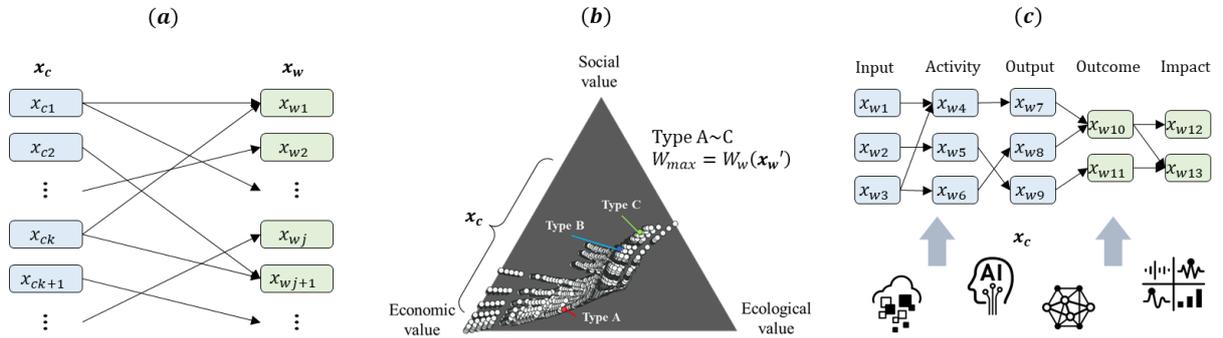

Figure 5. Policy evaluation method based on WE pluralism: (a) fact–value combined parameter system, (b) combined policy-making approach, and (c) combined impact evaluation.

In the fact–value combined parameter system shown in Figure 5 (a), the value parameter $x_w$ and the fact parameter $x_c$ that reflects it are connected by a network to infer the change in the value parameter $x_w$ in response to a policy, or manipulation of the fact parameter $x_c$ [11]. The value parameter $x_w$ is obtained by mapping $f^{-1}: X_n \to X_w$ from the subjective questionnaire (set $X_n$) to the aggregate results (set $X_w$) in response to Equations (9) and (10), and $x_w$ representing well-being $W = W_w(x_w)$ in response to Equation (11). It is safe to consider that the questionnaire implicitly assumes the consensus $W_n \circ f \equiv W_w$ that the aggregate results will be used.

The fact parameter $x_c$ is objective joint facts (set $X_c$) for joint fact-finding, and the parameter is chosen to be relational to the value parameter $x_w$. While in constitutive pluralism, the list of generic objective indicators is enumerated, resulting in a gap from the subjective indicators, the fact parameter $x_c$ used here reflects the context of WE in conceptual pluralism. There is a tradeoff between versatility and reflectivity, and $x_c$ here takes the latter position heavily. In other words, it is important how the fact parameter $x_c$ is set up to reflect the value parameter $x_w$. The policy is evaluated by $W' = W_w(x_w')$ using the relationship $x_w' = g(x_w, x_c)$ between the value parameter $x_w$ and the fact parameter $x_c$, corresponding to Equation (13), by performing the mapping $g: X_w \times X_c \to X_w'$.

In the combined policy-making approach shown in Figure 5 (b), the target function $W = W_w(x_w)$ representing well-being is formulated, the fact parameter $x_c$ representing the policy is combined with the target function, and the policy that maximizes the revised target function $W' = W_w(x_w')$ is selected [28]. The target function $W = W_w(x_w)$ has well-being as the target variable $W$ and factors such as social, environmental and economic as explanatory variables $x_w$, and it is obtained by multiple regression analysis of the aggregate results (set $X_w$) of the subjective questionnaire (set $X_n$) (corresponding to Equations (10) and (11): $f^{-1}: X_n \to X_w$ and $W_n \circ f \equiv W_w$).

The fact parameter $x_c$ is the result (set $X_c$) of the multi-agent simulation for the various policies' operating parameters. Depending on the number of combinations of operating parameters, tens of thousands of different results are obtained, as shown in the white circle plots on the ternary graph in Figure 5 (b). Then, by formulating the relational expression $x_w' = g(x_w, x_c)$ that combines these results $x_c$ and the explanatory variable $x_w$ in response to Equation (13), the policy that maximizes the revised target function $W' = W_w(x_w')$ is selected out of tens of thousands of possible results. The red (Type A), blue (Type B), and green (Type C) circles in Figure 5 (b) show that the desired policies differ depending on how the relational expression $x_w' = g(x_w, x_c)$ is formulated to express what balance of social, environmental, and economic values is to be emphasized.

In the combined impact evaluation shown in Figure 5 (c), a logic model $W = W_w(x_w)$ is established to achieve well-being, the fact parameter $x_c$ representing the policies is combined with the logic model, and the output of the logic model, impact $W' = W_w(x_w')$ is evaluated [29]. A logic model, as shown in the upper part of Figure 5 (c), generally describes and illustrates the cause-and-effect relationship of inputs (investments such as funds and personnel)→activities (organizational activities)→outputs (products and services)→outcomes (results produced)→impacts (social, environmental, and economic changes aimed at) [30]. Since the logic model is created by investors, local governments, nonprofit organizations, and corporations etc. under the agreement of stakeholders when they promote social businesses, it can be said to correspond to $W = W_w(x_w)$ in Equation (11).

For the fact parameters (set $X_c$) corresponding to the policy, it can be used that quantitative measurement data, forecast data based on statistical processing, and the results of multi-agent simulations similar to the combined policy-making approach in Figure 5 (b), and evaluate the impact $W' = W_w(x_w')$ of the policy corresponding to Equation (13) by establishing a relationship $x_w' = g(x_w, x_c)$ connecting these data $x_c$ and the variable $x_w$ in the logic model. Note that the fact parameter $x_c$ is often combined to the left side of the logic model, since logic models generally tend to become more subjective as one moves from left to right. In addition, depending on the policy, there may be a negative as well as positive impact, or positive or negative for multiple impacts, and the final evaluation of the policy is left to the stakeholders.

Through the above three examples, I was able to demonstrate the reality and usefulness of the policy evaluation methods presented in this paper. However, the setting of fact parameters that reflect value parameters in the fact–value combined parameter system, the coupling of explanatory variables of the target function and simulation results in the combined policy-making approach, and the coupling of the logic model and fact parameters in the combined impact evaluation are the keys to whether policies can be better evaluated. The techniques and skills involved are beyond the scope of this paper, but will be examined in future fieldwork in communities and municipalities.

## 4  Conclusion

In this paper, to address the issue of how to evaluate policies that contribute to well-being, I first started from subject–object pluralism and developed it into individual–group pluralism and WE pluralism. Furthermore, by incorporating the policy plurality into WE pluralism and combining it with joint fact-finding under the consensus of WE, a method for evaluating the effect of policies on well-being was presented. For reference, Figure 6 (a) and (b) again show schematically the policy evaluation methods based on subject (conceptual)–object (constitutive) pluralism and WE pluralism, respectively. In the former, it was difficult for objective indicators to reflect subjective context, and subjective indicators differed from individual to individual, while in the latter, WE reached a consensus (subjectively) on well-being, and then combined joint facts that WE could accept as possibly objective.

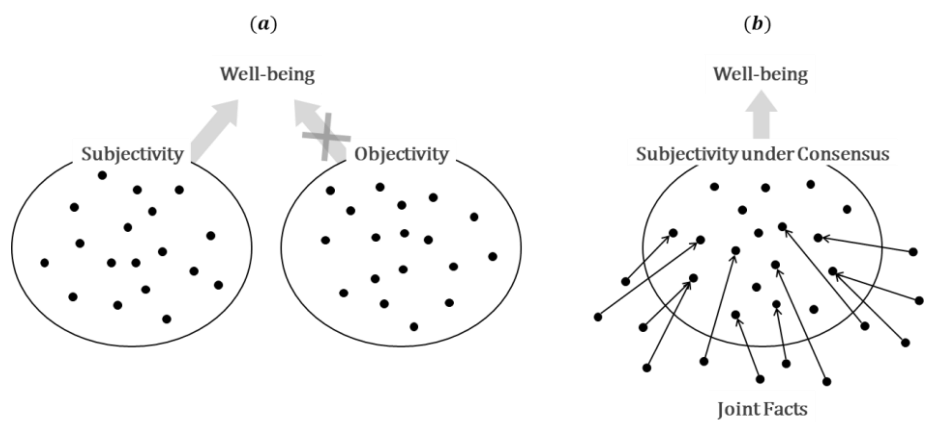

Figure 6. Comparison of policy evaluation methods: (a) subject (conceptual)–object (constitutive) pluralism, and (b) WE pluralism.

Finally, I will briefly discuss the significance of this paper and future issues. First, by developing from subject–object pluralism to individual–group pluralism and then to WE pluralism, and by including the context of conceptual pluralism in them, I avoided the problem of gap between subjective and objective indicators in constitutive pluralism. Second, instead of the dichotomy of individual subjectivity–objectivity, or individual subjectivity and group intersubjectivity, the "narrow-wide WE" concept was developed within the gradation of I–family–community–municipality–nation–world, a new groundwork for evaluating policies involving diverse stakeholders was provided. Third, I formulated a policy evaluation method that combines WE consensus and joint fact-finding, and demonstrated its usefulness by presenting examples of implementation. Future issues are to verify the policy evaluation method of this paper through fieldwork and to study the coupling method of consensus-based subjective value and joint fact-finding. However, the coupling method will be supported by statistical approaches such as behavioral economics and social psychology, and will again continue to pursue intersubjective and objective certainty.

## Additional Notes

For convenience, I summarize here a series of equations according to the synopsis of this paper.

| Argument | Equation | # |
|---|---|---|
| Subject–object pluralism | $W(x, p) = W_s(x, p) + W_o(x)$ | (1) |
| Individual–group pluralism | $W(x) = W_s(x, p) + W_{is}(x, g)$ | (2) |
| WE pluralism | $W(x) = W_n(x) + W_w(x)$ | (3) |
| Normalization | $W = r \cdot W_n(x_n) + (1 - r) \cdot W_w(x_w)$ | (7) |
| Policy plurality | $W = r \cdot W_n(\boldsymbol{x_n}) + (1 - r) \cdot W_w(\boldsymbol{x_w})$ <br> $X_n = \{x_{n1}, x_{n2}, \cdots, x_{ni}, \cdots\}$ <br> $X_w = \{x_{w1}, x_{w2}, \cdots, x_{wj}, \cdots\}$ | (9) |
| Mapping | $f^{-1}: X_n \to X_w$ <br> $\boldsymbol{x_n} = f(\boldsymbol{x_w})$ <br> $W = r \cdot W_n \circ f(\boldsymbol{x_w}) + (1 - r) \cdot W_w(\boldsymbol{x_w})$ | (10) |
| Consensus | $W_n \circ f \equiv W_w$ <br> $W = W_w(\boldsymbol{x_w})$ | (11) |
| Joint fact-finding | $X_c = \{x_{c1}, x_{c2}, \cdots, x_{ck}, \cdots\}$ | (12) |

| Mapping | $g: X_w \times X_c \to X_w{}'$ <br> $\boldsymbol{x_w}' = g(\boldsymbol{x_w}, \boldsymbol{x_c})$ | (13) |
|---|---|---|
| Policy evaluation | $W' = W_w(\boldsymbol{x_w}')$ | |

## Acknowledgement

In the project "Practical Examination of ELSI brought by Smartization of Community and Four-Dimensional Co-Creation Model" of JST "Responsible Innovation with Conscience and Agility," I would like to thank Professor Yasuo Deguchi of the Graduate School of Letters, Kyoto University for his guidance as principal investigator, Associate Professor Shunsuke Sugimoto of the Faculty of Business and Commerce, Keio University for his valuable and numerous suggestions on well-being and subjectivity as leader of the Mobile WE group, and Hitachi Kyoto University Laboratory for discussions in charge of the Community Evaluation Parameter Group. I also thank those who discussed this paper during the workshop at the 15th Annual Meeting of the Japan Association of Contemporary and Applied Philosophy. This work was supported by JST RISTEX Grant Number JPMJRS22J5, Japan.

## Author Information

Takeshi Kato

Hitachi Kyoto University Laboratory, Open Innovation Institute, Kyoto University, Kyoto, Japan

kato.takeshi.3u@kyoto-u.ac.jp